# OSCAR: A Collaborative Bandwidth Aggregation System

Karim Habak
School of Computer Science
College of Computing
Georgia Institute of Technology
Email: karim.habak@cc.gatech.edu

Khaled A. Harras
Computer Science Department
School of Computer Science
Carnegie Mellon University
Email: kharras@cs.cmu.edu

Moustafa Youssef
Department of CS. and Eng.
Egypt-Japan University of
Science and Technology (E-JUST)
Email: moustafa.youssef@ejust.edu.eg

*Abstract*—The exponential increase in mobile data demand, coupled with growing user expectation to be connected in all places at all times, have introduced novel challenges for researchers to address. Fortunately, the wide spread deployment of various network technologies and the increased adoption of multi-interface enabled devices have enabled researchers to develop solutions for those challenges. Such solutions aim to exploit available interfaces on such devices in both solitary and collaborative forms. These solutions, however, have faced a steep deployment barrier.

In this paper, we present OSCAR, a multi-objective, incentive-based, collaborative, and deployable bandwidth aggregation system. We present the OSCAR architecture that does not introduce any intermediate hardware nor require changes to current applications or legacy servers. The OSCAR architecture is designed to automatically estimate the system's context, dynamically schedule various connections and/or packets to different interfaces, be backwards compatible with the current Internet architecture, and provide the user with incentives for collaboration. We also formulate the OSCAR scheduler as a multi-objective, multi-modal scheduler that maximizes system throughput while minimizing energy consumption or financial cost. We evaluate OSCAR via implementation on Linux, as well as via simulation, and compare our results to the current optimal achievable throughput, cost, and energy consumption. Our evaluation shows that, in the throughput maximization mode, we provide up to 150% enhancement in throughput compared to current operating systems, without any changes to legacy servers. Moreover, this performance gain further increases with the availability of connection resume-supporting, or OSCAR-enabled servers, reaching the maximum achievable upper-bound throughput.

## I. Introduction

The Federal Communications Commission (FCC) has indicated an expected *data tsunami problem* predicting a 25-50× increase in mobile data traffic by the year 2015 [1], [2]. This expected explosive demand for mobile data, along with expensive data roaming charges and users' expectations to remain connected in all places at all time, are creating novel challenges for service providers and researchers to solve. The recent deployment and adoption of FON [3], a system that enables users to share their home bandwidth for obtaining WiFi coverage via other worldwide FON users, reflects a recent push for novel solutions to help solve a subset of these challenges. Other novel solutions are needed, however, when users are on the run or in public places like malls, airports, or bus and railway stations. Fortunately, the widespread deployment of various wireless technologies coupled with the exponential increase of multi-interface enabled devices are providing users with many alternatives for sending and receiving data.

A potential approach for solving the data tsunami problem as well as expensive data roaming charges for mobile users is exploiting all available communication interfaces available on modern mobile devices in both solitary and collaborative forms. In the solitary form, the goal is to exploit any direct Internet connectivity on any of the available interfaces by distributing applications data across them in order to achieve higher throughput, minimize energy consumption, and/or minimize cost. In the collaborative form, the goal is to enable and incentivize mobile devices to utilize their neighbors' under-utilized bandwidth in addition to their own direct Internet connections.

Utilizing the mobile devices interfaces in these two forms have been investigated over the past few years [4]. The focus has largely been on the solitary form developing a variety of multi-interface bandwidth aggregation solutions at different layers of the protocol stack [5]–[17]. Other attempts to utilize these interfaces in the collaborative form were also introduced [18]–[22]. These approaches either deal with a small scale collaborative community managed by a single authority [19], [20], or utilize proxy servers to handle and guarantee such collaboration [18], [21]. Overall, despite the fact that current smart phones, tablets, and other mobile devices are equipped with multiple network interfaces, current operating systems mainly allow users to utilize only one interface at a time, or enable limited bandwidth sharing options through tethering. In other words, there has been a high deployment barrier and solutions have focused on bandwidth maximization while not paying sufficient attention to energy and/or cost efficiency.

In this paper, we present OSCAR, a multi-objective, incentive-based, collaborative, and deployable bandwidth aggregation system. OSCAR fulfills the following requirements: (1) It is easily deployable without requiring changes to legacy servers, applications, or network infrastructure (i.e. adding new hardware like proxies and routers); (2) It is able to seamlessly exploit available network interfaces in solitary and collaborative forms; (3) It adapts to real-time Internet characteristics and the system parameters to achieve efficient utilization of these interfaces; (4) It is equipped with an incentive system that encourages users to share their bandwidth with others; (5) It adopts an optimal multi-objective and multi-modal scheduler that maximizes the overall system throughput, while minimizing cost and energy consumption based on the user's requirements and system's status; (6) It leverages incremental system adoption and deployment to further enhance performance gains.

Our contributions in this paper are: (1) Designing the OSCAR system architecture fulfilling the stated requirements above. (2) Formulating OSCAR's data scheduler as an optimal multi-objective, multi-modal scheduler that takes user requirements, device context information, as well as application requirements into consideration while distributing application data across multiple interfaces. (3) Developing OSCAR communication protocols in order to enable efficient and secured communication between the collaborating nodes as well as OSCAR enabled servers.

We evaluate OSCAR via implementation on Linux, as well as via simulation, and show its ability to increase the overall system throughput while achieving its cost and energy efficiency targets. The results show that, with no changes to current Internet architectures, OSCAR reaches the throughput upper-bound. It also provides up to 150% enhancement in throughput compared to the current Operating Systems without any change to legacy servers. Our results also demonstrate OSCAR's ability to maintain cost and the energy consumption levels in the user-defined acceptable ranges.

The remainder of this paper is organized as follows. Section II discusses the related work. A motivational scenario along with the intuition of our solutions is presented in Section III. We then present the overall architecture of our system in Section IV. Section V formulates the OSCAR scheduling problem, after which we present OSCAR's communication protocols in Section VI. In Section VII, we discuss our Linux OS implementation and evaluate OSCAR's performance and compare it to the current solutions. Finally, Section VIII concludes the paper and provides directions for future work.

## II. RELATED WORK

### A. Solitary Solutions

Many solutions have emerged to address single-device bandwidth aggregation at different layers of the protocol stack [4]. Application layer solutions either assume that applications are aware of the existing network interfaces and take the responsibility of utilizing them for their needs [12] or rely on modifying the kernel socket handling functions to enable existing applications use multiple interfaces [12], [13]. Such modifications require changes to legacy servers in order to support these new sockets. In addition, [13] requires feedback from the applications about their performance, and hence is not backwards compatible with previous versions of the applications.

Many bandwidth aggregation techniques, however, naturally lie in the transport layer [8]–[11]. These solutions replace TCP with mechanisms and protocols that handle multiple interfaces. Such techniques require changes to the client operating system, legacy servers, and/or applications; and hence have a high deployment barrier. Finally, the majority of the network layer solutions hide the variation in interfaces from the running TCP protocol [5]–[7]. Chebrolu et al. [5] require having a proxy server that communicates with the client and is aware of the client's multiple interfaces. Others implement their system at both connection-ends, which makes their deployment rely on updating the legacy servers [6]. On the other hand, MAR [7] requires having a special router as well as an optional proxy server.

Despite all this work, modern operating systems only allow users to use one of the available interfaces at a time, even if multiple of them are connected to the Internet. This attests to the fact that all current proposals for bandwidth aggregation face a steep deployment barrier. In addition, these solutions have largely focused only on maximizing throughput while overlooking other user and system goals like minimizing cost and energy consumption.

Recently, developing a deployable bandwidth aggregation system has been investigated [14], [15], [23]. DBAS focuses only on maximizing the overall system throughput [15], [23]. On the other hand, even though OPERETTA is deployable and energy-aware, it does not utilize the interfaces to their maximum in case of non-OPERETTA enabled servers [14]. In addition, similar to previous solutions, they only focus on utilizing the available interfaces in the solitary form, overlooking available connectivity at neighboring devices as well as cost efficiency.

### B. Collaborative Solutions

There exists some shy attempts to utilize the available interfaces in the collaborative form [18]–[22]. These solutions are motivated by the increase of the devices density and are aimed at sharing neighboring devices connectivity. Unfortunately, these systems have many shortcomings. Examples include requiring Internet infrastructure updates such as the existence of proxy servers [18], [19], [21], updating the available applications to interface with the new API and specify their requirements [20], developing applications that are fully responsible for handling and making use of the collaboration for their own benefits [22]. In addition, all these solutions focus only on maximizing throughput and usually ignore developing incentives systems that enable this collaboration.

The closest system to OSCAR in terms of incentivized bandwidth sharing is FON [3]. FON is a commercial system that enables users to share their home bandwidth for obtaining WiFi coverage via other worldwide FON users. Basically, each user installs a Fonera router in her home network. This Fonera router is responsible for sharing a portion of the user's bandwidth with other FON system users. The main differences between FON and OSCAR are: (1) FON does not aggregate the available bandwidth neither in a device nor in collaborative devices. (2) FON uses special router (Fonera). (3) FON routers are static while OSCAR targets mobile devices. (4) The user is unable to connect to more than one Fonera. (5) FON users are incetivized by the need of being connected to the Internet while being away from their home network. OSCAR builds on this need as well, but also adds a tangible financial pricing system to further encourage collaboration.

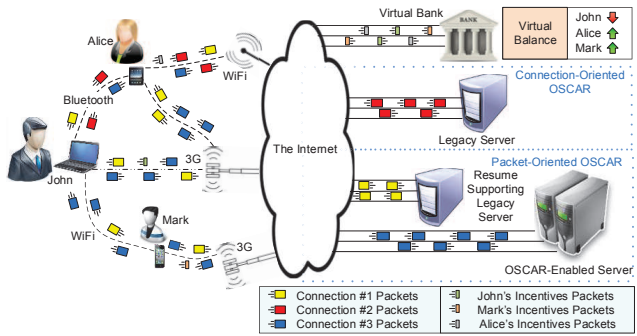

Fig. 1. OSCAR scenario

## III. MOTIVATING SCENARIO AND SOLUTION

Most mobile devices today are equipped with multiple heterogeneous network interfaces. For example, John's mobile device is equipped with Wifi, Bluetooth and 3G. When John is sitting in a mall, waiting for a train at the station, or having a meal, he typically watches youtube videos, listens to podcasts, uses facebook to get his social network feeds, and opens his favorite news cites. He connects his mobile device to available Wifi hotspots while being connected to his 3G network and leaving his Bluetooth idle. In such places John has a lot of people around him such as Mark who is not using his 3G Internet connectivity and Alice who is connected to a high bandwidth WiFi network but is not currently using it.

Unfortunately, John's current operating system assigns all his connections to Wifi. Consequently, John has to wait until the youtube video is buffered in order to watch it continuously without disruption; meanwhile other applications are slowly retrieving their content. With the high contention on the Wifi interface, the available bandwidth is degraded, and John disconnects his device from Wifi to utilize his 3G connection. Although, this increases his throughput, the available bandwidth is not sufficient to smoothly retrieve the content, he incurs the added overhead of restarting his non-resumable connections, and pays a higher cost in battery and money for using 3G.

John's needs can be satisfied by concurrently utilizing his available network interfaces as well as the under-utilized connections of Mark and Alice. John's applications traffic can be scheduled across different interfaces and neighbors in order to enhance his user experience. Such a scheduler should take the network interface characteristics, bandwidth heterogeneity, and different application traffic characteristics into account when scheduling different connections to the available interfaces. Furthermore, the scheduler should exploit servers support for the resume functionality in order to seamlessly migrate connections assigned to an interface that got disconnected to another. The scheduler can also leverage this resume functionality to schedule the data at a finer granularity in order to enhance the overall system performance. It would also be equipped with an incentive mechanism in order to convince John's neighbors to share their connectivity with him. Finally, this system would adopt an efficient and secured communication protocol to make the collaboration between the different users feasible.

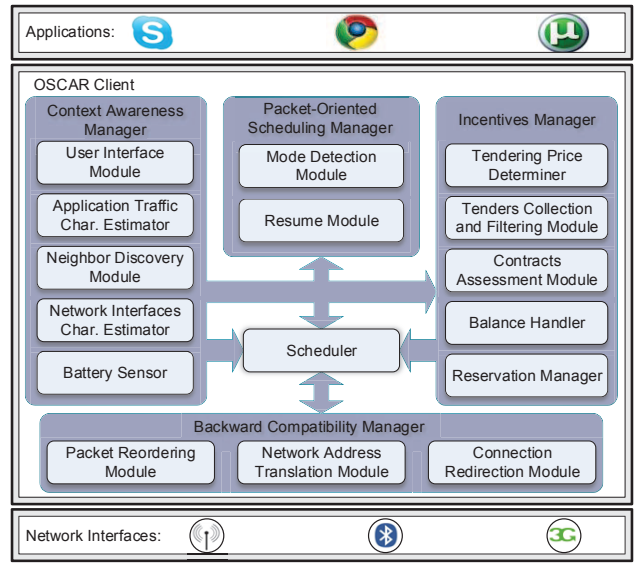

Fig. 2. OSCAR client architecture

Figure 1 shows a scenario, based on OSCAR, in which John's device is able utilize all the available connectivity options including Mark's and Alice's under-utilized interfaces to achieve John's requirements and enhances his experience. In this scenario, we notice that while communicating with the legacy servers the device schedules the connections to the paths in a connection-oriented mode, in which all the packets belonging to the same connection utilizes the same path. However, while being connected to an OSCAR enabled server or resume-supporting legacy server, it uses fine-grained packet-oriented scheduling for further performance enhancements. We also observe that there is also a virtual bank, which is part of the incentive system, that John uses to pay for the bandwidth he shared from Mark and Alice's. We now move on to discussing the architecture and system for enabling such scenario.

## IV. THE OSCAR SYSTEM

In this section, we start by describing our system assumptions. We then provide an overview of OSCAR's architecture followed by a detailed description of its main components.

### A. Assumptions

OSCAR works in a distributed environment, where a node can share and use the bandwidth available from its OSCAR-enabled neighbors to connect to both legacy and OSCAR-enabled servers. For this, we assume that all OSCAR components are trusted. Handling security aspects of the software modules is outside the scope of this paper. We further assume that there is a trusted entity, called the virtual bank, that is responsible for authenticating OSCAR users and keeping track of the virtual currency exchanged by users for the bandwidth (as detailed in sections IV-F and IV-I). Bandwidth sharing agreement between two nodes is performed in rounds, each with a fixed duration determined at the beginning of the round. To determine the cost of the bandwidth at the selling node, OSCAR has a module that suggests an initial cost based on different factors (as detailed in Section IV-F1). The user, however, can manually adjust the cost, e.g. to offer bandwidth sharing for her own devices for free. Once the price is fixed, it cannot be negotiated until the next round. The buyer collects the different offers from the neighboring buyers and selects the ones that best fits its needs. A buyer can re-sell her bandwidth to other nodes, allowing for multi-hop bandwidth sharing. OSCAR modules address the cost and looping issues associated with bandwidth re-selling.

### B. Communicating Entities

Figure 1 shows that we have five main communicating entities in our architecture. Firstly, **client devices** equipped with multiple network interfaces varying in their available bandwidth, energy consumption rates and cost per unit data. Each interface can be used to connect directly to the Internet and/or through neighboring client devices sharing their connectivity options. These options vary in their usage cost and available bandwidth to be shared. We call each of these possible ways to connect to the Internet a *path*. Each of these devices runs multiple applications that vary in their characteristics. Secondly, a **virtual bank** which is a trusted third party that handles the payment transactions between the different devices as well as the authentication related aspects. Thirdly, **legacy servers** which are typical non-modified Internet servers. The clients use the *connection-oriented* mode to connect with them as its default mode of operation. In this mode, OSCAR schedules different connections to the available paths such that a connection can be assigned to only one path. Once assigned, all the packets belonging to this connection utilize the same path. Fourthly, **connection resume-supporting legacy servers**, e.g. FTP and HTTP servers that support resuming the connections. OSCAR leverages these servers to enhance the performance by switching to a *packet-oriented mode*, where each packet or group of packets can be independently scheduled on a different path. Finally, **OSCAR-enabled servers**, running our software modules, which OSCAR leverages for enabling highly efficient *packet-oriented* scheduling.

### C. Architecture Overview

Figure 2 shows the OSCAR's **client** architecture. The architecture implements five core functionalities: First, context awareness (implemented through the Context Awareness Manager). It enables OSCAR to sense its environment; including user requirements, applications traffic characteristics, attached interfaces characteristics, battery status, and nearby OSCAR-enabled devices. Second, backward compatibility (implemented through the Backward Compatibility Manager); which enables OSCAR's clients to make use of the available connectivity opportunities without requiring any changes to legacy servers, network infrastructure, or legacy applications. Third, leveraging connection resume-supporting legacy servers and OSCAR-enabled servers for further performance enhancements (implemented through the Packet-Oriented Scheduling Manager). Fourth, Handling the user incentives and collaborating with the trusted bank module in handling the payment transactions implemented through the (Incentive Manager). Finally, optimally scheduling the connections and/or packets on the different paths in order to achieve the user requirements (implemented in the OSCAR Scheduler).

Note that, OSCAR server architecture is a subset of this client architecture containing the following modules: (1) Mode Detection Module, (2) Packet Reordering Module, (3) Scheduler, (4) Network Interfaces Characteristics Estimator, and (5) Network Address Translation Module. For the balance of this section, we describe each of these components in more details.

### D. Context Awareness Manager

One of the key factors that enables OSCAR to efficiently utilize all available connectivity options is its ability to determine, store, and utilize the mobile device's context information accurately and efficiently. This is the responsibility of the context-awareness manager. This component consists of the following five modules: the User Interface Module, the Application Traffic Characteristics Estimator, the Neighbor Discovery Module, the Network Interface Characteristics Estimator, and the Battery Sensor.

*1) User Interface Module:* OSCAR provides a user-friendly interface that enables users to set their requirements, preferences, and scheduling policies. This also enables users to input other contextual parameters like the interface usage cost (e.g. cost of using 3G or the hotel Wifi). Finally, this module is also used to express willingness to share bandwidth, set the amount to be shared, and define the needed revenue in return for consuming the device's battery to aid others.

*2) Application Traffic Characteristics Estimator:* Knowing the application traffic characteristics significantly impacts OSCAR's performance. However, to be fully deployable, we can not require any changes to existing applications in order to determine this crucial information. Our approach is to automatically estimate application characteristics based on their operational history on given port numbers. These estimation techniques enable us to accurately determine the application traffic characteristics since most popular applications have specific ports reserved for their communication and exhibit relatively consistent behavior. We estimate the average connection data demand in bytes for each port as follows:

After a connection is terminated, the module updates the estimated values of connection data demand ($C_{\text{demand}}$) as:

$$C_{\text{demand}}(i) = (1-\rho)C_{\text{demand}}(i-1) + \rho TC_{\text{demand}}(i) \quad (1)$$

where $TC_{\text{demand}}(i)$ is the number of bytes transmitted by the $i^{\text{th}}$ connection that uses this port number which has just finished, $C_{\text{demand}}(i)$ is the new estimate of the connection data demand, and $\rho$ is a smoothing coefficient, taken equal to 0.125 to evaluate the equation efficiently. This way, the estimate can be obtained by an efficient shift operation.

TABLE I
POWER CONSUMPTION TABLE [25]

| Interface | Tech. | Δ Power[1] | Data rate |
|---|---|---|---|
| Netgear MA701 | Wifi | 726 mW | 11Mbps |
| Linksys WCF12 | Wifi | 634 mW | 11Mbps |
| BlueCore3 | Bluetooth | 95 mW | 723.2Kbps |

*3) Neighbor Discovery Module:* This module is responsible for discovering the existence of nearby OSCAR-enabled devices. It selects the interfaces that are going to be used in the discovery process, determines the frequency of sending neighbor discovery queries, maintains the lists of the discovered neighbors, and communicates with the virtual bank to authenticate these neighbors.

*4) Network Interface Characteristics Estimator:* This module is responsible for estimating the characteristics of each network interface. In particular, it estimates the available bandwidth, energy consumption rates, the used physical data rates, and the achievable maximum data rate for each interface. To estimate the available bandwidth, it periodically connects through each interface to various geographically dispersed servers that support the packet-pair estimation technique [24]. On the other hand, when communicating with OSCAR-enabled servers, the module periodically sends the data packets in pairs and marks them in order to be used for estimating the available bandwidth using the packet-pair technique. These estimates are sufficient since bandwidth bottlenecks typically exist at the client's end not the server's, which is typically connected to high bandwidth links and designed to scale with the number of clients.

To characterize the energy consumption, Table I shows how the power consumption of a network interface depends on the NIC, the technology used, and the physical transmission data rate [25]. Hence, in order to make OSCAR estimate the energy consumption for its interfaces, we build an online service with a database containing energy consumption rates for various network interfaces at different physical transmission data rates. Once OSCAR runs, it queries the database for the energy consumption rates of each network interface attached to the device.

Estimating the physical transmission data rate is the simplest task since it can be queried using the supported OS API. It is used in addition to the queried energy consumption rates to determine the interface's current energy consumption rate. We use the estimated physical transmission rate as an estimate for the maximum achievable data rate for the interface.

*5) Battery Sensor:* This module senses whether the device is plugged to a power source or running on battery along with the available battery level.

### E. Packet-Oriented Scheduling Manager

Utilizing the opportunities of packet-oriented scheduling is one of the key features introduced by OSCAR. Detecting such opportunities is the responsibility of the packet-oriented scheduling manager. This module exploits the availability of OSCAR-enabled servers or servers that support the connection resumption to enhance the performance of OSCAR. It contains two sub-modules: the mode detection module and the resume module.

*1) Mode Detection Module:* This module detects whether the end point of the connection supports OSCAR or not in order to enable the optional packet-oriented mode. It is implemented as a service that monitors the exchanged data packets between the two end points and alters the TCP packet header. When the client starts a new connection, the module alters the TCP header adding OSCAR existence flag in the option part of the TCP header, if the server responds similarly, this indicates that OSCAR is supported on the server. Hence, packet-oriented mode can be utilized. If so, OSCAR will schedule the packets that belong to this TCP connection on the available paths. If OSCAR is not supported on the server, the client resorts to the default connection-oriented mode unless the connection is resumable.

This approach enables OSCAR to efficiently determine the operation mode without incurring extra packet transmission overhead. It also enables us to avoid extra delay since immediate scheduling in connection-oriented mode occurs until we determine the need to switch to packet-oriented mode.

*2) Resume Module:* This module is utilized when communicating with legacy servers, with the purpose of enabling packet-oriented scheduling even if the server is not OSCAR-enabled. To detect whether a particular server supports the resume functionality or not, we build a web service maintaining a list of protocols that support the resume functionality. OSCAR periodically connects to this web service to update and cache this list of protocols. Each protocol is represented by a template that uniquely identifies the protocol, e.g. its port number, as well as how to check with the end server that it supports the resume functionality, e.g. checking for the "Accept-Ranges" header option in HTTP. The OSCAR Resume Module uses such protocol information to determine whether the server supports the resume functionality or not, and accordingly enables the optional packet-oriented mode.

The resume module is implemented as an in-device proxy responsible for identifying the resumable connections, scheduling sub-connections across different device interfaces, and closing these sub-connections once they terminate. To minimize the latency and buffer size the application needs to wait for the in-order delivery of data, the module uses two approaches: (1) it divides the connection into fixed length units, where a unit with an earlier offset from the beginning of the connection is retrieved over the multiple interfaces before the next unit is retrieved. Each interface is assigned a portion of the unit to retrieve based on weights decided by the scheduler. (2) It sorts the interfaces in a descending order according to their scheduling weight and starts the sub-connection with the larger weight first. This way, larger amounts of data will arrive in order from the beginning of the connection. For further optimization, if the resumable connection supports range requests, it does not close the sub-connection once the load is finished but reuses it for the next load assigned to that interface. This optimization technique avoids the extra delay of increasing the congestion windows to its optimal size introduced by TCP slow start.

On the other hand, this module also handles interface disconnection which increases in the collaborative mode due to device mobility. It does so by monitoring each sub-connection ensuring it will receive its portion. If one or more of the sub-connections' paths became unavailable because of disconnection, it re-schedules this sub-connection on one of the available running paths. The scheduler is responsible for determining the appropriate path to be used in this case.

### F. Incentives Manager

The success of a collaborative bandwidth aggregation system is tightly coupled with its ability to encourage users to collaborate and share their available bandwidths. Incentive systems have been widely studied in literature for a wide array of applications where users share resources; approaches proposed generally fall into (1) reputation-based approaches and (2) credit-based approaches.

In reputation based incentive approaches [26], [27], nodes earn a reputation based on their collaboration with their neighbors who then spread this reputation accordingly. Hence, the nodes carry the overhead of monitoring their neighbors and spread their observations in the network to enable other nodes to determine their reputation level and act accordingly. These approaches, however, are typically suitable for systems with stable memberships that last long enough to earn high reputations. They are not suitable for deployable collaborative bandwidth aggregation systems, where nodes will randomly meet at random locations at any point in time (e.g. passengers at an airport or a train station).

Credit-based systems are more suitable for such networks since they usually rely on a trusted third party that maintains credit for the communicating nodes [28], [29]. Like free markets, nodes in such systems collaborate with each other and pay for the services acquired from one another.

The Incentives Manager in our architecture is designed to enable bandwidth sharing amongst users by adopting a credit-based system. It is particularly responsible for handling virtual currency exchange and bandwidth reservation. These functions are implemented using the following five modules: the Tendering Price Determiner, the Tenders Collection and Filtering Module, the Contracts Assessment Module, the Balance Handler, and the Reservation Manager.

*1) Tendering Price Determiner:* This module runs at the *seller* and is responsible for determining a price to offer for sharing a device's bandwidth on a given interface for a fixed amount of time. It proposes an initial price covering the cost for using the shared bandwidth at interface $j$ (which is connected to the Internet) while communicating with a buyer through the seller interface $i$ based on the environmental status[2] and other system parameters. The device owner can agree with this initial price suggested by the system or preemptively change it through the User Interface Module. To determine an initial price, multiple cost parameters are taken into account including the service provider cost, the energy consumption cost, and the market status. This cost is fixed for a fixed amount of time and needs to be renewed in order to handle mobility and disconnections of devices.

**a. Service Provider Cost:** Since a node may be paying for its own Internet access (either through an Internet provider or a neighboring sharing node), this node needs to at least cover such cost in order to avoid incurring a loss. Hence, the first portion of the incentives price will be $c_i$ which is the service provider cost per unit data of using interface $i$.

**b. Energy Consumption:** Sharing bandwidth by enabling multiple network interfaces will consume extra power, which is an extremely crucial resource especially for battery operated mobile devices. Hence, a device sharing its bandwidth needs some compensation for its consumed energy. Noting that a selling device uses two of its interfaces (one connected to the Internet, $i$, and the other connected to the buyer, $j$), calculating the required compensation should take into account the following parameters: (1) the energy consumed to relay a unit of data for interfaces $i$ and $j$ would equal $(e_i + e_j)$, (2) the battery capacity ($E_{\text{Capacity}}$), and (3) the remaining battery ratio ($R_{\text{remaining}}$). Hence we calculate the energy compensation factor ($\text{EC}_{ij}$) as follows:

$$\text{EC}_{ij} = [\gamma + (1 - R_{\text{remaining}})] \frac{e_i + e_j}{E_{\text{Capacity}}} \quad (2)$$

Where $\gamma$ is a binary value reflecting whether the device is connected to a power source or not. If connected, $\gamma = 0$, the user still needs to be compensated for the consumed energy as part of the charging power is still used in the bandwidth sharing. Therefore, the importance of your energy consumption degrades till it reaches 0 when the battery is fully charged. On the other hand, if the device is running on battery power, $\gamma = 1$ guarantees receiving appropriate compensation even if the battery is fully charged.

**c. The Market Status:** With potentially multiple devices requesting and offering bandwidth, the demand and supply mini-market status amongst these devices is a crucial pricing factor. A balance is required in order to maximize the sharing node's benefit while providing priority to users willing to pay more for important data they need to transmit. We therefore calculate the market dependent cost (MC) as follows:

$$\text{MC}_{n,i} = \max \left[ \text{MC}_{n-1,i} + \left( \frac{(1+\kappa)B_{i,\text{Requested}} - B_{i,\text{Offered}}}{B_{i,\text{Offered}}} \right) \eta, 0 \right] \quad (3)$$

Where $i$ is the index of the interface connected to the Internet that we share its bandwidth, $\text{MC}_{n-1,i}$ is the market status cost during the previous estimation session, $B_{i,\text{Requested}}$ is the sum of the requested bandwidth for reservation during the last session from such interface (i.e. demand), $B_{i,\text{Offered}}$ is the offered bandwidth from that interface (i.e. supply), $\eta$ is the cost increase factor determined by the user, and $\kappa$ is a revenue multiplier. The reason behind using the revenue multiplier ($\kappa$) is to allow for increasing the price when the supply is equal to the demand and is set to a small value (0.1 in our case).

---
[2]Environmental status is the current market status including requests on the seller and reservations.

The intuition is that as long as the demand is higher than the supply, we keep increasing the market status cost value. This increase is a function of the cost variation factor ($\eta$) as well as the normalized difference between the demand and supply.

**The Incentive Price:** After discussing the various factors that should be taken into consideration while calculating the required incentives price we put them together in the following equation:

$$\begin{aligned} I_{n,i,j} &= \nu \text{EC}_{i,j} + \text{MC}_{n,i} + c_i \\ &= \nu \left[ \gamma + (1 - R_{\text{remaining}}) \right] \frac{e_i + e_j}{E_{\text{Capacity}}} + c_i + \\ &\quad \left[ \text{MC}_{n-1,i} + \left( \frac{(1+\kappa) B_{i,\text{Requested}} - B_{i,\text{Offered}}}{B_{i,\text{Offered}}} \right) \eta \right]^+ \end{aligned}$$
(4)

Where $\nu$ is an energy-cost conversion factor that maps the energy cost ($\text{EC}_{i,j}$) to monetary cost. The value of $\nu$ is set by the virtual bank based on the current fair market prices.

At the end, this incentive price is proposed to the user and she can freely change it seeking more or less profit. Although setting $I_{n,i,j} > c_j$ can be used to avoid loss and loops, a user has the ability to set the price even to zero to satisfy cases when she shares bandwidth with other devices belonging to her or to family members. The IP of the interface with the farthest OSCAR device in the chain is always used in the tenders to detect and avoid loops.

Once this cost is determined and sent to the buyer, the buyer either accepts or rejects it.

*2) Tenders Collection and Filtering Module:* This module runs at the bandwidth requestor and is responsible for collecting tenders from nearby OSCAR enabled devices. These tenders contain the path available bandwidth and the price of transmitting a unit of data in this path (as calculated by the Tendering Price Determiner at the seller). After successfully collecting the tenders, the module sorts them based on cost and forwards the least cost ones whose aggregate bandwidth saturates the requester bandwidth to the scheduler. This way, the scheduler can take the available bandwidth through neighbors in its decision.

*3) Contracts Assessment Module:* When two neighbors agree on the set of terms for bandwidth sharing (i.e. cost, reserved bandwidth, and duration), both the buyer and seller monitor the traffic to regulate and ensure these terms. Note that this is enforced by the trusted OSCAR components. This module is also responsible for renewing the contract before it expires.

*4) Balance Handler:* This module is responsible for handling the payment transactions with the virtual bank. It also makes sure that buyer node does not make requests that exceed its available credit.

*5) Reservation Manager:* This module is designed to keep track of the reserved bandwidth on each node to avoid using an interface above its capacity. It builds a reservation table that contains the neighbor identifier, the reserved bandwidth, and time stamp for the most recent usage. A reservation time out ($RT_{\text{Out}}$) is used to avoid indefinite reservation that may occur due to mobility or the existence of a malicious node.

### G. Backward Compatibility Manager

In order to make OSCAR deployable in today's Internet, it should be backwards compatible with ***both the current legacy servers and applications***. This is the responsibility of the backward compatibility manager that consists of the Packet Reordering Module, the NAT Module, and the Connection Redirection Module.

*1) Packet Reordering Module:* This module is activated for the packet-oriented mode to handle packet reordering issues introduced by transmitting data using multiple network interfaces to be compatible with legacy applications. It delays the packets as well as their acknowledgments before passing them to the upper layer. This is designed to avoid the performance degradation introduced by using TCP over multiple paths. In order to avoid being over-protective, it maintains an estimation of the TCP's RTT in order to forward the out-of-order data to TCP before the timeout to prevent excessive drops in the congestion window.

*2) Network Address Translation Module:* When a connection go through a sharing neighbor to a legacy server, since a legacy server can see only one IP address for the client, OSCAR needs to change the IP address in the packet to the neighboring client IP. Otherwise, the reply from the legacy server will go directly to the client, rather than to the sharing neighbor, removing the benefit of sharing the bandwidth with neighbors.

To address this, we implement a Network Address Translation (NAT) module at each node which is activated on the sharing (seller) node. This module does the NATing operation for the sent packets and reverses this operation for the received packets from the legacy server.

*3) Connection Redirection Module:* This module is responsible for redirecting the newly requested connections to the the Resume Module in order to enable utilizing the connection resumes while hiding this from legacy applications.

### H. Scheduler

This is the heart of the OSCAR system. It takes its input from all the previously mentioned modules in order to schedule the connections and/or packets to the different available paths. We discuss the OSCAR scheduler in details in Section V.

## I. Virtual Banks

This is a trusted third party that maintains user accounts and credit balances. It is also responsible for handling the payment process between the different users as well as OSCAR's authentication.

## V. OSCAR SCHEDULER

In this section, we formulate the optimal OSCAR scheduling problem. We begin by describing our system model followed by the optimal scheduling problem.

### A. System Model

Table II summarizes the system parameters and notations. We assume a mobile device with $m$ different paths to the Internet. Each path represents a way to connect to the Internet either by using the Interface's direct connectivity or by using the interface to communicate with one of its neighbors sharing its Internet connectivity. Each of those paths has its effective bandwidth $b_j$ and cost per unit data $c_j$, which can be the service provider usage cost or the cost paid to the neighbor in order to use their connectivity. In addition, each path uses one network interface and it has an energy consumption rate $a_j$, where $a_j$ equals the difference in power consumption between the active and idle states of the used interface. The data rate of each path interface is denoted as $r_j$. The device runs a set of connections that share these interfaces and varies in their characteristics.

Our scheduling unit is a connection or a packet. We refer to a standard network connection as a ***stream*** to avoid confusion with the scheduling unit. Scheduling decisions are taken when a new stream (number of streams active in the system is $n$, including the new stream) is requested from an application. The Mode Detection Module (Section IV-E1) and the Resume Module (Section ) then determine whether the operation mode is connection-based ($S_n = 1$), or packet-based ($S_n = 0$) when the other end is OSCAR-enabled or supports the resume mode. In the former case, the scheduler's goal is to determine to which path it should be assigned (sets $x_{nj} = 1$ for only one path $j$). In either case, the percentage of packets to be assigned to each path, i.e. paths relative packet load ($w_j$), should be re-calculated based on the current system load ($\mathcal{L}$). Our OSCAR scheduler has three modes of operation based on user preferences: (1) throughput maximization, (2) energy minimization, and (3) cost minimization.

### B. Optimal Scheduling

In this section, we formulate our scheduler and its different modes of operation. Generally, the decision variables are: (1) If $S_j = 1$, which path to assign the new stream $n$ to (variable $x_{nj}$) and (2) the new values for $w_j$, $\forall j : 1 \leq j \leq m$.

We start by presenting the formal definition of the Throughput Maximization Mode (Section V-B1). Then, we present the formal definition of the Energy Minimization Mode (Section V-B2). After that, we present the formal definition of the Cost Minimization Mode (Section V-B3) followed by the common constraints that must be satisfied regardless of operation

TABLE II
LIST OF SYMBOLS USED

| Symbol | Description |
|---|---|
| $\mathcal{T}$ | The overall system throughput |
| $\mathcal{L}$ | The current system load |
| $\mathcal{L}_i$ | The current system load for stream $i$ |
| $\mathcal{S}_i$ | Determines whether stream $i$ is connection-based (1) or packet-based (0) |
| $b_j$ | The effective bandwidth of path $j$ |
| $r_j$ | The data rate of path $j$ |
| $a_j$ | Difference in power between active and idle states of path $j$ |
| $\mathcal{E}$ | The energy consumed in order to transfer the system load |
| $\mathcal{E}_{\text{avg}}$ | The average energy consumed per unit data while transferring the system load |
| $\mathcal{E}_j$ | The energy consumed for path $j$ to transfer its load |
| $c_j$ | The cost per unit data of path $j$ |
| $\mathcal{C}$ | The system load transferring cost |
| $\mathcal{C}_{\text{avg}}$ | The average cost per unit data of transmitting the system load |
| $\mathcal{C}_j$ | Path $j$ cost of transferring its load |
| $\Delta_j$ | Path $j$ needed time for finishing its load |
| $x_{ij}$ | For connection-oriented streams, equals 1 if stream $i$ is assigned to interface $j$. Equals 0 otherwise. |
| $w_j$ | The ratio of packets assigned to interface $j$ |
| $n$ | Number of active streams including the new request |
| $m$ | Number of different paths |

mode (Section V-B4). Finally, we present our problem solution (Section V-B5).

*1) Throughput Maximization Mode:* In this mode, the scheduler's goal is to maximize the system throughput under certain energy and cost constraints. In particular, the user puts a limit on the average cost per unit data ($\mathcal{C}_{\text{avg,Target}}$) as well as a limit on the energy consumption per unit data ($\mathcal{E}_{\text{avg,Target}}$) that should not be exceeded.

**Objective Function:**

The objective of the scheduler at any decision instance is to maximize the overall system throughput ($\mathcal{T}$). Given the system load ($\mathcal{L}$), the objective can be written as:

$$\text{Maximize } \mathcal{T} = \frac{\mathcal{L}}{\max_j \Delta_j} \quad (5)$$

where $\Delta_j$ is the time needed for path $j$ to finish all its load (connection and packet based). Since $\mathcal{L}$ is constant, the objective function is equivalent to:

$$\text{Minimize } \max_j \Delta_j$$

$$= \text{Minimize } \max_j \left( \frac{\sum_{i=1}^{n} \left( \mathcal{L}_i \left(1 - \mathcal{S}_i\right) w_j \right) + \sum_{i=1}^{n} \left( \mathcal{L}_i \mathcal{S}_i x_{ij} \right)}{b_j} \right) \quad (6)$$

Where the left summation represents the packet-oriented mode load (each term, $\mathcal{L}_i \left(1 - \mathcal{S}_i\right) w_j$, is the number of bytes from the packet-oriented stream $i$ load assigned to path $j$) and the right summation is the connection-oriented mode load. Note that any stream $i$ will be either connection-oriented ($\mathcal{S}_i = 1$) or packet-oriented ($\mathcal{S}_i = 0$) and thus will appear in only one of the two summations. Dividing the sum of two loads by the available bandwidth on that path ($b_j$) gives the time needed for path $j$ to finish its load.

**Constraints:**

The following constraints must be satisfied:

***Target Cost***: As we mentioned, the user puts a limit ($\mathcal{C}_{\text{avg,Target}}$) on the average cost she is willing to pay per unit data.

$$\mathcal{C}_{\text{avg}} \leq \mathcal{C}_{\text{avg,Target}} \quad (7)$$

Since

$$\mathcal{C}_{\text{avg}} = \frac{\sum_{j=1}^{m} c_j \left[\sum_{i=1}^{n}\left(\mathcal{L}_i\left(1-\mathcal{S}_i\right)w_j\right) + \sum_{i=1}^{n} \mathcal{L}_i \mathcal{S}_i x_{ij}\right]}{\mathcal{L}} \quad (8)$$

Substituting in (7):

$$\sum_{j=1}^{m} c_j \left(w_j \sum_{i=1}^{n}\left(\mathcal{L}_i\left(1-\mathcal{S}_i\right)\right) + \sum_{i=1}^{n} \mathcal{L}_i \mathcal{S}_i x_{ij}\right) \leq \mathcal{L} \mathcal{C}_{\text{avg,Target}} \quad (9)$$

Where $c_j$ is the average cost per unit data while using path $j$. $\sum_{i=1}^{n}\left(\mathcal{L}_i\left(1-\mathcal{S}_i\right)w_j\right) + \sum_{i=1}^{n} \mathcal{L}_i \mathcal{S}_i x_{ij}$ is the load assigned to path $j$ as in (6).

***Target Energy***: Similarly, the client's device can tolerate up to a certain level of the average energy consumed per unit data.

$$\mathcal{E}_{\text{avg}} \leq \mathcal{E}_{\text{avg,Target}} \quad (10)$$

Since

$$\mathcal{E}_{\text{avg}} = \frac{\sum_{j=1}^{m} \frac{a_j}{r_j} \left[\sum_{i=1}^{n}\left(\mathcal{L}_i\left(1-\mathcal{S}_i\right)w_j\right) + \sum_{i=1}^{n} \mathcal{L}_i \mathcal{S}_i x_{ij}\right]}{\mathcal{L}} \quad (11)$$

Substituting in (10):

$$\sum_{j=1}^{m} \frac{a_j}{r_j} \left(\sum_{i=1}^{n}\left(\mathcal{L}_i\left(1-\mathcal{S}_i\right)w_j\right) + \sum_{i=1}^{n} \mathcal{L}_i \mathcal{S}_i x_{ij}\right) \leq \mathcal{L} \mathcal{E}_{\text{avg,Target}} \quad (12)$$

Where $\frac{a_j}{r_j}$ is the average energy consumption per unit data while using path $j$.

*2) Energy Minimization Mode:* In this mode, the scheduler's goal is to minimizing the overall energy consumed under certain throughput and cost constraints. In particular, the user puts a limit on the average cost per unit data to $\mathcal{C}_{\text{avg,Target}}$ and requests minimum throughput of $\mathcal{T}_{\text{target}}$

**Objective Function:**

The objective of the scheduler at any decision instance is to minimize the overall system energy consumption ($\mathcal{E}$). This objective can be written as:

$$\text{Minimized } \mathcal{E} = \sum_j \mathcal{E}_j$$

where, $\mathcal{E}_j$ is the energy consumption of path $j$. This energy consumption can be divided into two parts: (1) energy needed to finish the load of the connection-oriented streams and (2) energy needed to finish the load of the packet-oriented streams. The former is equal to $(a_j/r_j) \sum_{i=1}^{n} \mathcal{L}_i \mathcal{S}_i x_{ij}$ where $a_j/r_j$ is the energy consumed per unit data while using path $j$ and $\sum_{i=1}^{n-1} \mathcal{L}_i \mathcal{S}_i x_{ij}$ is the connection oriented load assigned to path $j$. Similarly, the later is equal to $(a_j/r_j) \sum_{i=1}^{n} \mathcal{L}_i\left(1-\mathcal{S}_i\right)w_j$. Since all the connection-oriented streams but the newly arriving one cannot be re-assigned to other interfaces, the objective function becomes:

$$\text{Minimize } \mathcal{E} = \sum_{j=1}^{m} \frac{a_j}{r_j} \left(w_j \sum_{i=1}^{n}\left(\mathcal{L}_i\left(1-\mathcal{S}_i\right)\right) + \mathcal{L}_n \mathcal{S}_n x_{nj}\right) \quad (13)$$

Where: $\frac{a_j}{r_j}$ refers to the energy consumed per unit data. $w_j \sum_{i=1}^{n}\left(\mathcal{L}_i\left(1-\mathcal{S}_i\right)\right) + \mathcal{L}_n \mathcal{S}_n x_{nj}$ the load assigned to interface $j$ except the previously assigned connection oriented load $(a_j/r_j) \sum_{i=1}^{n-1} \mathcal{L}_i \mathcal{S}_i x_{ij}$. We removed this part from the equation since it is constant and will not affect the minimization process.

**Constraints:**

The following constraints must be satisfied:

***Target Throughput***: As we mentioned, in this mode, the client's device has to achieve at least a certain level of throughput ($\mathcal{T}_{\text{Target}}$). Therefore,

$$\mathcal{T} = \frac{\mathcal{L}}{\max_j \Delta_j} \geq \mathcal{T}_{\text{Target}} \quad (14)$$

where $\Delta_j$ is the time needed for path $j$ to finish all its load ($\mathcal{L}$) (connection and packet based). The throughput is the ratio between the load on the system and the time needed to transfer this load.

$$\rightarrow \forall j, \Delta_j = \frac{\sum_{i=1}^{n}\left(\mathcal{L}_i\left(1-\mathcal{S}_i\right)w_j\right) + \sum_{i=1}^{n}\left(\mathcal{L}_i \mathcal{S}_i x_{ij}\right)}{b_j} \leq \frac{\mathcal{L}}{\mathcal{T}_{\text{Target}}}$$

Rearranging:

$$\rightarrow \forall j, w_j \sum_{i=1}^{n}\left(\mathcal{L}_i\left(1-\mathcal{S}_i\right)\right) + \mathcal{L}_n \mathcal{S}_n x_{nj} \leq \frac{\mathcal{L} b_j}{\mathcal{T}_{\text{Target}}} - \sum_{i=1}^{n-1}\left(\mathcal{L}_i \mathcal{S}_i x_{ij}\right) \quad (15)$$

Note that the RHS is constant.

***Target Cost***: Similar to the throughput maximization mode (Section V-B1), the target cost constraint is represented by Equation 9

*3) Cost Minimization Mode:* In this case, the user is interested in achieving at least a certain level of throughput $\mathcal{T}_{\text{target}}$ with limitations on the average energy consumed per unit data $\mathcal{E}_{\text{avg,target}}$

**Objective Function:**

The overall objective of the scheduler is to minimize the overall system cost needed ($\mathcal{C}$) to consume the system load:

$$\text{Minimized } \mathcal{C} = \sum_j \mathcal{C}_j$$

where, $\mathcal{C}_j$ is the cost needed for path $j$. This cost can be divided into two parts: (1) cost needed to finish the load of the connection-oriented streams and (2) energy needed to finish the load of the packet-oriented streams. The former is equal to $c_j \sum_{i=1}^{n} \mathcal{L}_i \mathcal{S}_i x_i j$, where $c_j$ is the cost per unit data for path $j$ and $\sum_{i=1}^{n} \mathcal{L}_i \mathcal{S}_i x_i j$ the amount of data assigned to this path in connection-oriented mode. Similarly, the later is equal to $c_j \sum_{i=1}^{n} \mathcal{L}_i\left(1-\mathcal{S}_i\right)w_j$. Since connection-oriented streams cannot be re-assigned to other paths, the objective function becomes:

$$\text{Minimize } \mathcal{C} = \sum_{j=1}^{m} c_j \left( w_j \sum_{i=1}^{n} (\mathcal{L}_i (1 - \mathcal{S}_i)) + \mathcal{L}_n \mathcal{S}_n x_{nj} \right) \quad (16)$$

**Constraints:**

The following constraints must be satisfied:

*Target Throughput*: Similar to the throughput maximization mode (Section V-B2), the target cost constraint is represented by Equation 15

*Target Energy*: Similar to the throughput maximization mode (Section V-B1), the target cost constraint is represented by Equation 12

*4) Common Constraints:* Here, we present our set of constrains that must be satisfied regardless of the scheduling mode.

*Integral Association*: If the new stream is connection-oriented, it should be assigned to only one path:

$$\sum_{j=1}^{m} x_{nj} + (1 - S_n) = 1 \quad (17)$$

Note that when $S_n = 0$, $x_{nj} = 0, \forall j$, which is the case when the new stream is determined to be packet-oriented by the Mode Detection Module.

*Packet Load Distribution*: For packet-oriented streams, their total load should be distributed over all interfaces:

$$\sum_{j=1}^{m} w_j = 1 \quad (18)$$

*Variable Ranges*: The trivial constraints for the range of the decision variables

$$1 \geq w_j \geq 0, 1 \leq j \leq m \quad (19)$$

$$x_{nj} \in \{0, 1\}, 1 \leq j \leq m \quad (20)$$

*5) Solution:* In general, this problem is a mixed 0-1 Integer Programming problem, which is an NP-complete problem. However, it has a special structure that allows for an efficient solution. In particular, we have two cases: if the new stream that triggered the scheduling decisions is packet-based ($S_n = 0$) and if it is connection-based ($S_n = 1$).

**Solution for the packet-based case:**

In this case, $x_{nj} = 0 \, \forall j$. The problem becomes a standard linear programming problem. Hence, it can be solved efficiently to determine the different values of $w_j$.

**Solution for the connection-based case:**

In this case, we need to determine the binary variables $\forall_j x_{nj}$ such that only one of them equals 1 and others are 0. Our algorithm sets each one of them to 1 and then solves the resulting linear programming problem to find $\forall_j w_j$. The value that achieve the best objective is then selected as the optimal decision.

### C. Discussion

In some cases, the selected constraints by the user may lead to an infeasible solution. For example, it may be the case that there is no assignment strategy that achieves both the cost and energy consumption constraints specified by the user concurrently. To eliminate this possibility and make the selection process user friendly, we design an iterative selection policy where the user selects her top priority constraint first (e.g. limit on cost). Based on this selection, OSCAR calculates the feasible range for the second constraint (e.g. energy consumption) and the user is asked to choose a value from this range.

## VI. OSCAR COMMUNICATION PROTOCOLS

In order to implement OSCAR, we developed two simple protocols. First, a collaboration protocol that handles bandwidth sharing amongst collaborating nodes. Second, an OSCAR server communication protocol for detecting OSCAR-enabled servers and exchanging control information between the client and server. In this section we give an overview on our OSCAR communication protocols followed by a detailed description for each protocol.

### A. Protocols Overview

In order to enable the collaboration between nodes, the OSCAR collaboration protocol aims to: (1) efficiently discover the neighbors, (2) authenticate both buyers and sellers, (3) enable cost agreement and (5) and guarantee the payment.

On the other hand, the OSCAR Server communication protocol aims to (1) discover the existence of OSCAR-enabled servers, (2) enable the server to know the list of (IP, port) pairs that they can use to send their data to the clients, (3) and exchange with the clients the packet scheduling weights. We give the details of both protocols in the next two subsections.

### B. OSCAR Collaboration Protocol

Figure 3 shows the state diagram of this protocol. The communicating entities in this protocol are client devices offering or requesting bandwidth (separated by vertical line) and the trusted bank server supporting our incentive system. The figure also shows the states generally broken down to the following three phases (separated by the horizontal lines): discovery and authentication, tendering, and payment.

### C. Discovery and Authentication

In order to maintain energy-efficient neighbor discovery, a requesting node sends neighbor discovery requests only when it has active packet-oriented connections or new connection requests. For this discovery, we adopt the eDiscovery [30] protocol and apply it to the available interfaces to determine the best discovery duty cycle for each interface.

Once eDiscovery fires a neighbor discovery event, our OSCAR protocol engages and broadcasts a Discovery Request (DReq) packet. This discovery packet has the header depicted in Figure 4(a). It contains the OSCAR version running at the requesting client, header length, the discovery request header data.

If several offering/selling neighbors exist, a requesting node receives discovery request (DReq) packets from each one. Once the requesting node reaches the discovery timeout, it parses all DRes replies, obtains all the user ids, then authenticates these respondents if it has their public keys. If at least one of the respondents is unknown to the requester, It starts the *bank-based neighbor authentication*.

*Bank-based Neighbor Authentication:* This authentication occurs when an OSCAR client (offering or requesting) needs to authenticate one or more of its neighbors. An authentication request (AReq) with headers shown in Figure 4(b) is sent to the bank module. The AReq packet has an operation code reflecting the discovery operation, the requesting user id to be used with the signature in authentication and integrity check, the number of requested entities to be authenticated, a list of requested user ids to be authenticated, a private key-based signature of the packet header data. Once the bank receives this packet, it authenticates the sender and checks the message integrity. Upon success, an authentication response (ARes) packet is sent back to the requesting client containing the public keys for the requested accounts, all signed with the requesting client private key.

### D. Tendering

This stage begins after a requesting node has received and authenticated DRes replies from neighbors. The node then broadcasts a tenders gathering request (GReq) packet using the header and information depicted in Figure 4(a) with the appropriate operation code. When an offering neighbor receives a GReq, it calculates the required tender for each of its connectivity options and sends a tenders packet. In addition to the usual packet information, the offering neighbor utilizes the operation related options to include the tender informaion. Each tender record contains a tender index, cost of unit data, the available bandwidth for sharing, and the tender duration. The requesting node receives multiple tenders from the neighboring offering nodes. Once the requesting node reaches the gathering timeout, it sends the tenders to the core OSCAR client which filters and then supplies them to the scheduler. The scheduler's responsibility at this time is to determine the accepted tenders based on the amount of extra bandwidth required and their cost. Once decided, the offering node sends a bandwidth reservation request (RReq) that contains in the optional part: the index of the tender selected as well as the amount of bandwidth needed for reservation. Finally, the offering node responds with a reservation acknowledgement (RACK) in case of successfully reserving the bandwidth or a negative acknowledgment (RNACK) when the requested bandwidth in not available (ex. reserved for someone else).

### E. Payment

After each service period, the offering node prepares a receipt for what has been consumed by the requesting node and signs the hashed version of this receipt with its private key. Once the requesting node receives that receipt, it hashes operation code, the requesting clients user id in order to be used for authentication and a private key-based signature of the packet header data. This signature signs a hashed version of the packet header with padding.

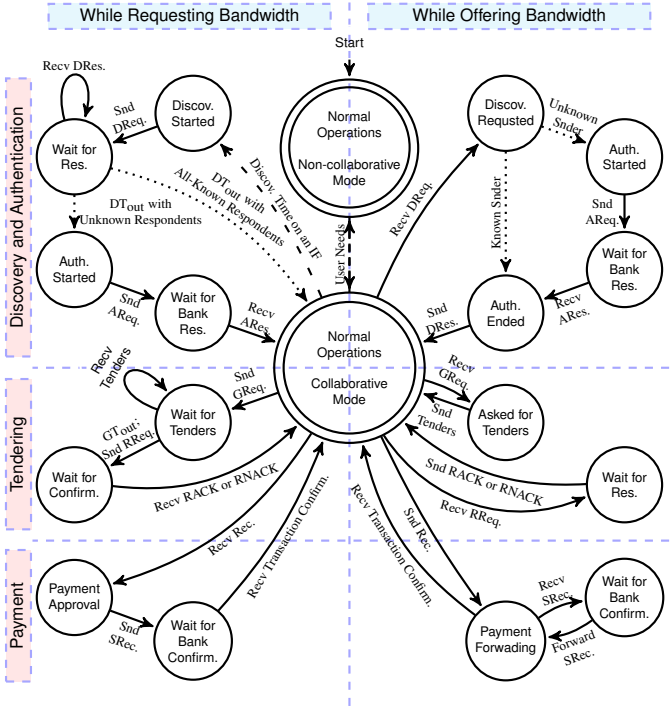

Fig. 3. OSCAR collaboration protocol state diagram.

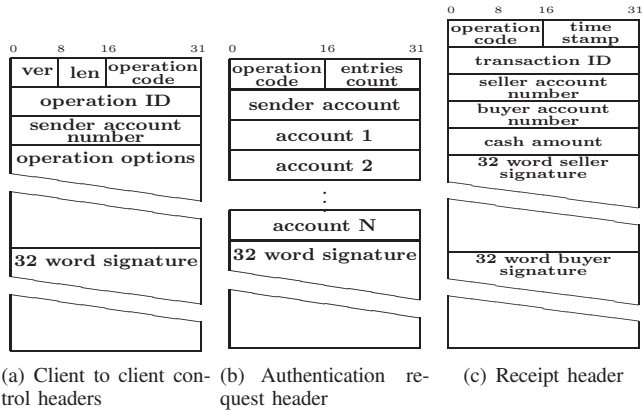

(a) Client to client control headers  (b) Authentication request header  (c) Receipt header

Fig. 4. OSCAR collaboration protocol headers

When a node offering bandwidth receives a discovery request (DReq) packet, it checks if it knows the public key of the requesting node based on its user id included in the request message. If so, it can authenticate that neighbor and verify the requested packet's integrity. Otherwise, it starts the *bank-based neighbor authentication*. Once the requester is authenticated, the receiving node sends a Discovery Response (DRes) packet. This packet has the same DReq header shown in Figure 4(a) and contains the highest version of OSCAR supported at both the requesting and the offering nodes, header length, the discovery response operation code, the offering node's user id, and a private key-based signature of the packet this packet and signs the hashed version with its private key. If the requesting node has a direct connection to the bank module, it sends the signed receipt to it. Otherwise, it forwards the signed receipt to its neighbor in order for it to be forwarded to the bank. The receipt follows the header format depicted in Figure 4(c). Both parties wait for the bank confirmation and the commitment of the transaction prior to further collaborations.

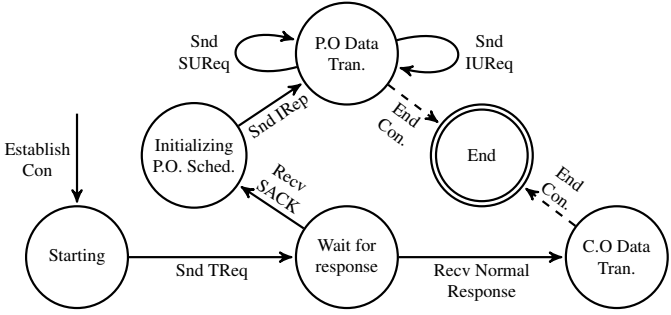

Fig. 5. OSCAR server communication protocol state diagram

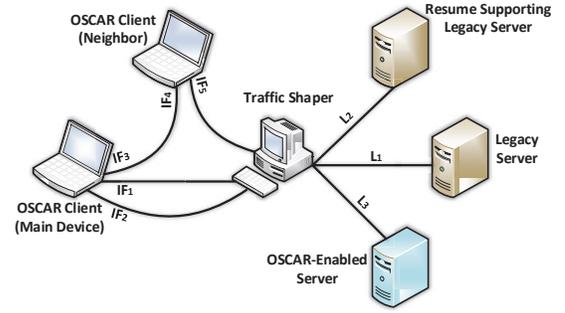

Fig. 6. Experimental setup.

### F. OSCAR Server Communication Protocol

The objective of this protocol is to enable the detection of OSCAR enabled servers for optimal resource utilization. Figure 5 shows this process that begins with the establishment of a new connection. An OSCAR client embeds the connection establishment exchanged packets with a server type inquiry request (TReq). It puts OSCAR's identification code in TCP's options header. If the server response contains the same OSCAR identification code, this is considered a server acknowledgement (SACK) from an OSCAR-enabled server, which allows the client to switch to the packet-oriented scheduling mode. Otherwise, the server will send a normal response (i.e. no TCP options used) which indicates that the server is not OSCAR-enabled. Hence, the client operates in the connection-oriented mode.

When the client operates in the packet-oriented mode, it sends identification reports (IRep) through all its interfaces to the server to identify its different IPs. Afterwards, packet-oriented data transmission begins. While utilizing the available paths to server, the client scheduler may change the scheduling weights. It sends a scheduling update request to the server (SUReq) in order to adapt to these changes. If new connections or interfaces are available (e.g. the user turned 3G on or entered a Wifi area), the client sends interface update request packet (IUReq) in order to handle these aspects.

## VII. SYSTEM IMPLEMENTATION AND PERFORMANCE EVALUATION

In this section we evaluate the performance of OSCAR via implementation. We begin with an overview of the implementation. Afterwards, we describe the experimental setup followed by the evaluation results for each scheduling mode. All the results have also been validated using NS2 simulations.

### A. Implementation

In addition to implementing the OSCAR communication protocols, incentive system, and scheduler as described in the previous sections, in this section, we briefly discuss extra details regarding our OSCAR Linux implementation. We specifically highlight the network layer middleware, application layer resume service, and the monitoring application we developed.

*1) OSCAR Network Layer Middleware:* We implement the OSCAR Network Layer Middleware using the Click modular router [31]. This approach allows the user to install OSCAR without recompiling the Linux kernel, and to intercept packets to and form the network protocol stack. It includes all the OSCAR modules except the User Interface Module and Resume Module (Section IV). Each component is implemented as a Click element responsible for processing the exchanged packets.

*2) Application Layer Resume Service:* This service reflects the Resume Module (Resume Module (Section )). It cooperates with the OSCAR middleware in order to make use of the legacy servers resume support. In particular, OSCAR directs all application connections and their data to this service, which analyzes the connection parameters, enables the packet-oriented scheduling mode if the connection is resumable, and issues the resume requests accordingly.

On the other hand, when OSCAR is initially installed, it connects to an online service which we provide that contains the templates of all the available resumable protocols. Each protocol template uniquely identifies its protocol, how to check with the end server that it supports the resume functionality, and how to perform resume requests. Periodically, the resume service connects to the online running service in order to receive the template of any newly appearing protocols.

*3) Monitoring Application:* The monitoring application represents the User Interface Module that captures the user's preferences, interface usage policies, and further monitors OSCAR behavior. It allows the user to select her average cost and energy consumption limitations, monitor her achieved throughput level, and set her interface usage costs and energy consumption preferences. It also provides the ability to perform manual assignment of certain applications categories (e.g. realtime applications) to certain interfaces (e.g. Ethernet). Finally, for testing purposes, this module allows the user to monitors OSCAR's internal data structures and estimated values by interfacing with the OSCAR middleware.

TABLE III
EXPERIMENTAL INTERFACES CHARACTERISTICS.

| Network Interface | Power (mWatt) | Cost ($/Mb) | D. Rate (Mbps) | BW (Mbps) |
|---|---|---|---|---|
| $IF_1$ (Wifi) | 634 | 0 | 11 | 1 |
| $IF_2$ (3G) | 900 | 0.02 | 42 | 2 |
| $IF_3$ (Blue.) | 95 | 0 | 0.7232 | 0 |
| $IF_4$ (Blue.) | 95 | 0 | 0.7232 | 0 |
| $IF_5$ (Wifi) | 726 | 0 | 11 | 1 |

TABLE IV
EXPERIMENTS PARAMETERS

| Parameter | Range | Nominal |
|---|---|---|
| $\forall_i L_i$ Bandwidth (Mbps) | 6 | 6 |
| $IF_1$ Bandwidth (Mbps) | 0.25 - 2 | 1 |
| Incentive Cost ($/Mb) | 0.03 | 0.03 |
| Neighbor sharing ratio (%) | 0 - 100 | 100 |

### B. Experimental Setup

Figure 6, Table III and Table IV depict our testbed and summarize the main parameters and values adopted in our evaluation. Our testbed consists of six nodes: an OSCAR-enabled server, two legacy servers with only one of them supporting the resume functionality, a main client, a neighboring device sharing its bandwidth, and a traffic shaper. The traffic shaper runs the NIST-NET [32] network emulator to emulate the varying network characteristics of each interface. Both clients are enabled with multiple network interfaces. On the main client, we run different applications that vary in terms of the number of connections per second they open ($\beta$), the average connection data demand ($\lambda$) and their destination port numbers. The client is connected to the traffic shaper node through two interfaces: $IF_1$ and $IF_2$. It is connected to the neighboring device via $IF_3$ at the client and $IF_4$ at the neighbor. The neighboring device is connected to the traffic shaper through $IF_5$. Each server is connected to the traffic shaper using a single high bandwidth link ($L$). We note that the combined bandwidth of $IF_1$, $IF_2$ and $IF_5$ is less than each server bandwidth in order to test the true impact of varying the interface characteristics and scheduling strategies. We define $\gamma \in [0, 100]$ as the percentage of connections that have the OSCAR-enabled servers as their destination. When $\gamma = 0$, all connections are with legacy servers, when $\gamma = 100$ all the connections are with OSCAR enabled servers.

We evaluate OSCAR using two classes of applications: browsers and FTP applications. Browsers have an average connection data demand of $\lambda_{\text{HTTP}} = 22.38KB$ [33]. FTP applications have an average data demand of $\lambda_{\text{FTP}} = 0.9498MB$ [33]. The connection establishment rate follows a Poisson process with mean ($\beta$) connections per second ($\beta_{\text{HTTP}} = 13\ con/sec$, and $\beta_{\text{FTP}} = 1\ con/sec$). Each experiment represents the average of 15 runs. Note that since OSCAR estimates the application characteristics, its performance is not sensitive to the specific application characteristics.

### C. Throughput Maximization Mode Results

In this section we evaluate the performance of OSCAR's Throughput Maximization Mode using three metrics: throughput, energy consumption per unit data, and cost per unit data. We vary the following parameters: the percentage of connections with OSCAR-enabled servers ($\gamma$), the percentage of resumable connections ($\alpha$), interface characteristics, neighbor sharing ratio, application workloads, connection heterogeneity, and Resume Module division unit. We compare the OSCAR optimal scheduler to three baseline schedulers:

- **Throughput Upper Bound**: This scheduler represents the theoretically maximum achievable throughput.
- **Round Robin (RR):** which assigns streams or packets to network interfaces in a rotating basis.
- **Weighted Round Robin (WRR):** Similar to the RR scheduler but weighs each interface by its estimated bandwidth; interfaces with larger bandwidths have proportionally more packets or streams assigned to them.

*1) Effect of Changing Streams with OSCAR-Enabled Servers ($\gamma$) vs Resumable Streams ($\alpha$):* As discussed in Section IV, OSCAR leverages both resume-supporting legacy servers and OSCAR-enabled servers to enhance performance. Figure 7 shows the effect of increasing the percentage of streams established with OSCAR-enabled servers ($\gamma$) on the performance of the OSCAR scheduler for different values of $\alpha$ (the percentage of resumable streams to legacy servers). In this experiment we set the energy consumption and cost limits to their maximum limits (131.36 Joule/Mb and 0.03 $/Mb respectively) to highlight the throughput gains that can be achieved by OSCAR. Based on Figure 7(a), we share the following observations: (1) Even when $\gamma = 0$ and $\alpha = 0$ (i.e. only working with legacy servers with **no** resume support), OSCAR can enhance the throughput by 150% as compared to the current OSs. (2) When $\gamma$ and $\alpha$ are low, most of the streams are connection-oriented, rendering the scheduling decision coarse grained; once the stream is assigned to a path, all its packets have to go through this path until termination. This reduces the scheduling optimality. (3) For $\alpha = 0\%$, the system reaches its throughput upper bound when we have only 30% of the streams connecting to OSCAR-enabled server ($\gamma = 30$). (4) This need of OSCAR-enabled servers decreases as $\alpha$ increases, till it reaches 0% when $\alpha = 35\%$, which is typical in the current Internet [34]. (5) The performance gain before saturation by adding more OSCAR-enabled servers is better than adding more legacy servers with resume support. We believe that this is due to the the resuming process overhead.

Figure 7(b) shows that OSCAR significant increase in throughput comes with even better cost for the user. This can be explained by noting that the current OSs use the interface with the maximum throughput, which happens to be the costly 3G interface in our case. OSCAR, on the other hand, mixes the different interfaces, leading to lower cost.

Figure 7(c) shows, as we relaxed the constraints on the energy consumption, that OSCAR uses more energy consump-tion than the current operating systems to achieve its superior throughput gains. This energy consumption can be further reduced by the user if needed by setting a limit on energy consumption.

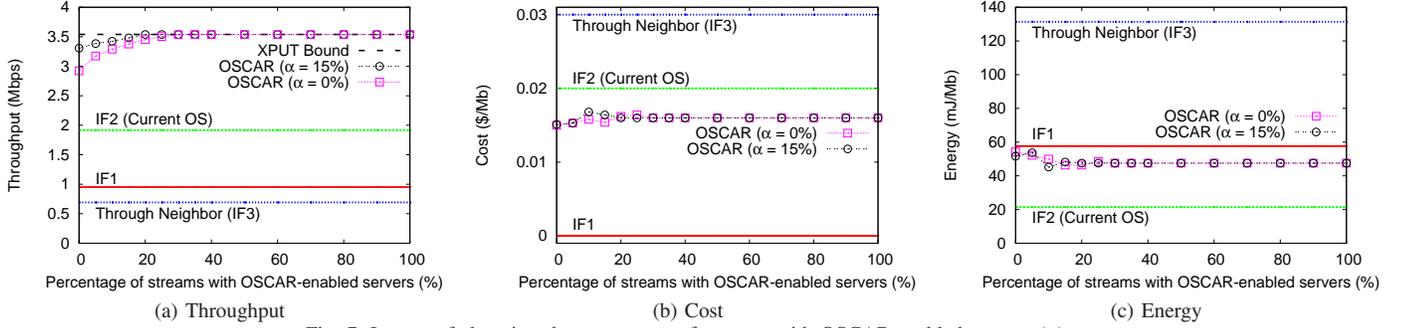

Fig. 7. Impact of changing the percentage of streams with OSCAR-enabled servers ($\gamma$).

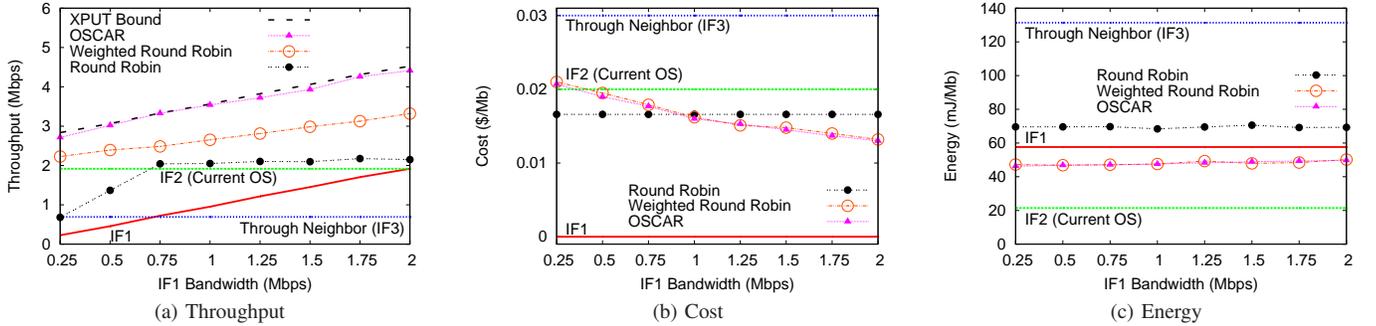

Fig. 8. Impact of interface heterogeneity.

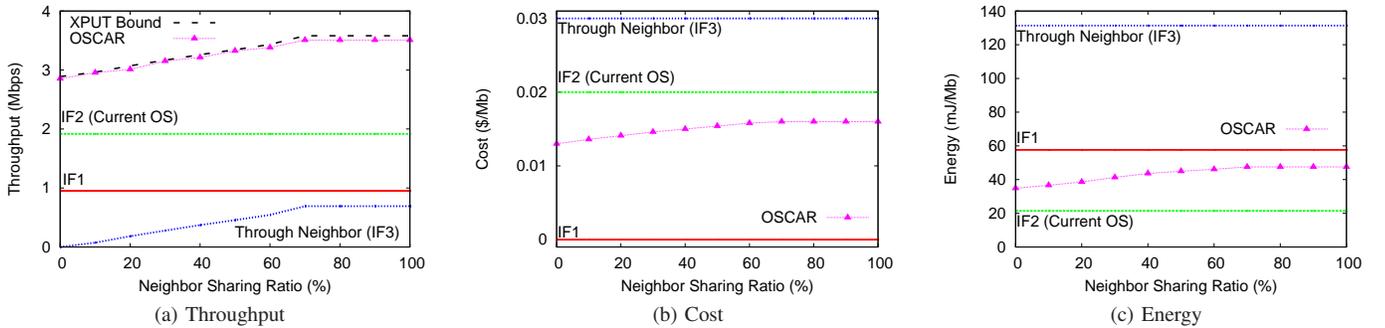

Fig. 9. Effect of changing the neighbor sharing ratio.

*2) Impact of Interface Heterogeneity:* In this experiment, we change the bandwidth of $IF_1$ from 0.25 Mbps to 2 Mbps while fixing the other parameters. Here we compare the OSCAR scheduler to the round robin (RR) and the weighted round robin (WRR) schedulers. The energy and cost constraints are still relaxed.

Figure 8(a) shows that when the bandwidth of $IF_1$ is low, using only $IF_2$ (i.e. the current OSs approach) outperforms the round robin scheduler. This is due to the fact that the round robin scheduler does not take the interface characteristics into account, assigning streams to each network interface in turn. Therefore, the low bandwidth interface becomes the bottleneck. We also notice that using the OSCAR scheduler outperforms the weighted round robin scheduler since our model takes the application characteristics into account while scheduling the connections as well as it makes use of the resume support in the legacy servers.

Figure 8(c) shows that for the WRR and OSCAR schedulers, the energy consumption per unit data slightly increases as the bandwidth of $IF_1$ increases. This is due to leveraging the increased bandwidth for higher throughput, though with the higher energy consumption of interface $IF_1$. The RR scheduler, on the other hand, is not affected by the increased bandwidth of $IF_1$ as its assignment policy is not bandwidth sensitive.

Similar behavior is observed in Figure 8(b), which shows that the average cost per unit data, noting that $IF_1$ has the lowest cost.

*3) Changing the Neighbor Sharing Ratio:* Figure 9 shows the effect of changing the neighbor sharing ratio. The figure shows that OSCAR can dynamically leverage the extra bandwidth available from the neighbor. The saturation at the sharing ratio of 70% is due to reaching the limit of the local interface ($IF_3$), even though the neighbor connection to the Internet ($IF_5$) can support a higher bandwidth. This highlights

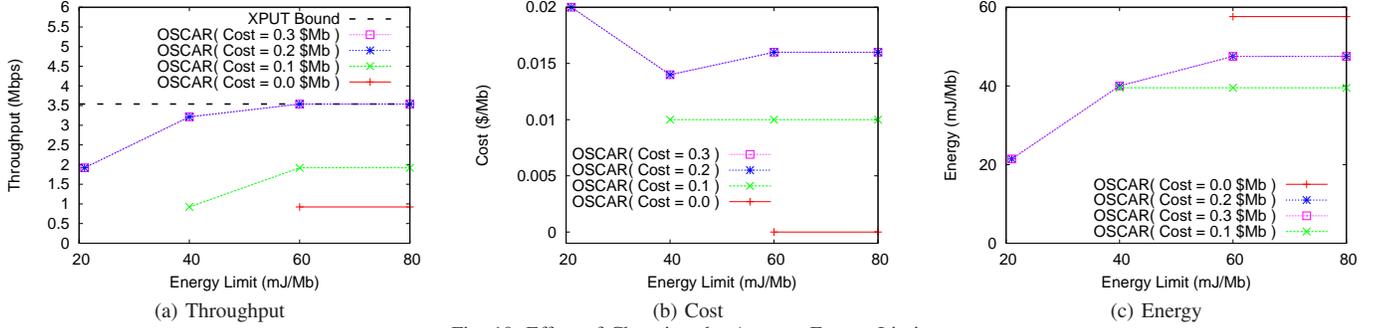
Fig. 10. Effect of Changing the Average Energy Limit

the OSCAR can estimate the bottleneck accurately.

*4) Changing the Cost and Energy Constraints:* Figure 10 shows the effect of changing the average energy consumption and cost limits on the performance of the OSCAR scheduler. In this experiment, we first select the energy consumption constraint and then the cost constraint. Figure 10(a) shows that, as expected, relaxing either the energy or cost constraint leads to better throughput. In addition, a more stringent energy constraint leads to a smaller feasibility region for the cost. This explains the discontinuity in the figure. OSCAR handles this transparently in a user friendly manner as discussed in Section V-C.

Another interesting notice in Figure 10(c) is that relaxing the cost constraint does not always lead to reducing the energy consumption and vice versa; *Since the scheduler goal in this evaluation is to maximize the throughput*, relaxing the cost constraint can lead to higher energy consumption if this leads to increasing the throughput while not violating the energy constraint.

For example, Figure 10(b) shows that, for a highly relaxed cost constraint (cost limit $\geq 0.02$), increasing the energy limits initially reduces the average cost incurred. However, the average cost incurred starts to increase after the cost energy limit becomes more relaxed at 40 mJ/Mb. For this particular case, initially, with a strict energy constraint, the expensive $IF_2$ was the only option since it has the minimum energy consumption. When the cost energy limit became 40 mJ/Mb, OSCAR started to utilize the other less expensive but energy consuming interface ($IF_1$), which leads to better throughput. Further relaxation of the energy consumption constraint led OSCAR to start using the neighboring device, which explains the increase of the average cost.

### D. Energy Minimization Mode Results

In this section we evaluate the performance of OSCAR's Energy Minimization Mode using three metrics: throughput, energy consumption per unit data, and cost per unit data. Simply, if we relaxed the cost and the throughput constraints, OSCAR's Energy Minimization Mode utilizes only the minimum energy consuming path. Therefore, to evaluate the performance of this mode, we vary the minimum throughput requirement and average cost limit.

Figure 11 shows the effect of changing the minimum required throughput and the average cost limit on the performance of OSCAR scheduler. In this experiment we first select the throughput constraint and then the cost constraint. Figure 11(c) shows that, as expected, relaxing either cost or throughput constraints leads to less energy consumption per unit data. In the extreme, when the cost limit is relatively high and the required throughput is very low, OSCAR decides to assign all the system traffic to the path with the minimum energy consuming per unit data which uses $IF_2$. In addition, with the increase of the required throughput or the decrease of the cost limit, the energy consumption per unit data increases. Note that the higher the throughput requirement may lead to smaller feasibility region for the cost. This explains the discontinuity in the figure. OSCAR handles this transparently in a user friendly manner as discussed in Section V-C.

Figure 11(a) shows OSCAR's ability to meet the user throughput requirement. Note that when the minimum energy requirement is very low, OSCAR achieve significantly higher throughput than required because it assigned all the traffic to the path with the minimum energy consumption per unit data and it did not have the incentive to limit its bandwidth because both the cost of unit data and energy consumption per unit data are constant.

Figure 11(b) shows that increasing the throughput requirement does not always have the same effect on the cost per unit data. We notice that initially increasing the throughput requirement did not affect the cost since OSCAR used the same path for transmitting the whole traffic. Therefore, the cost per unit data was constant when $\mathcal{T}_{target} \leq 2$Mbps. When the target throughput ($\mathcal{T}_{target}$) became greater than 2Mbps, OSCAR gradually started utilizing interface $IF_1$ cost-free path. Hence, the average cost per unit data started decreasing till the target throughput ($\mathcal{T}_{target}$) reached 3Mbps. When the target throughput ($\mathcal{T}_{target}$) exceeded 3Mbps, OSCAR had to start using the expensive and energy hungry path through $IF_3$ to achieve the required throughput. Hence, the cost started to increase with the increased user throughput requirement.

### E. Cost Minimization Mode Results

In this section we evaluate the performance of OSCAR's Cost Minimization Mode using three metrics: throughput, energy consumption per unit data, and cost per unit data. Sim-ilarly, if we relaxed the energy and the throughput constraints, OSCAR's Cost Minimization Mode utilizes only the most cost-efficient path. Therefore, to evaluate the performance of this mode, we vary the user throughput requirement and average energy limit.

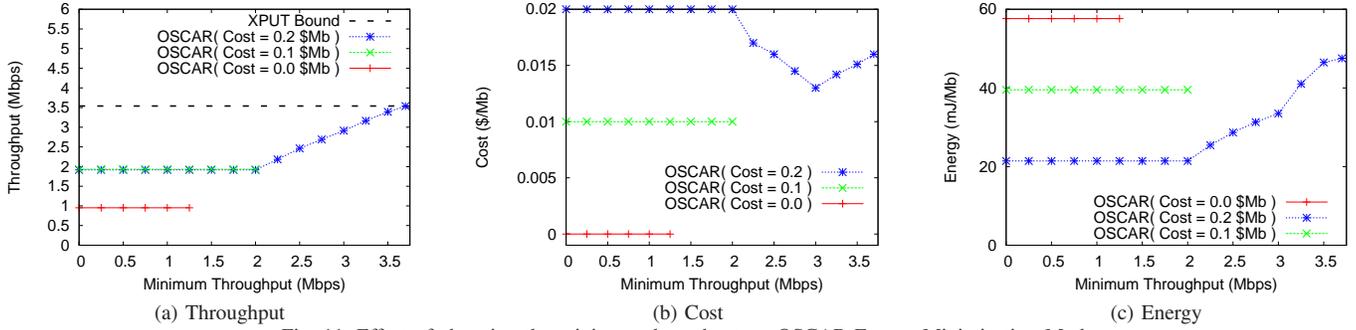

(a) Throughput  (b) Cost  (c) Energy

Fig. 11. Effect of changing the minimum throughput on OSCAR Energy Minimization Mode

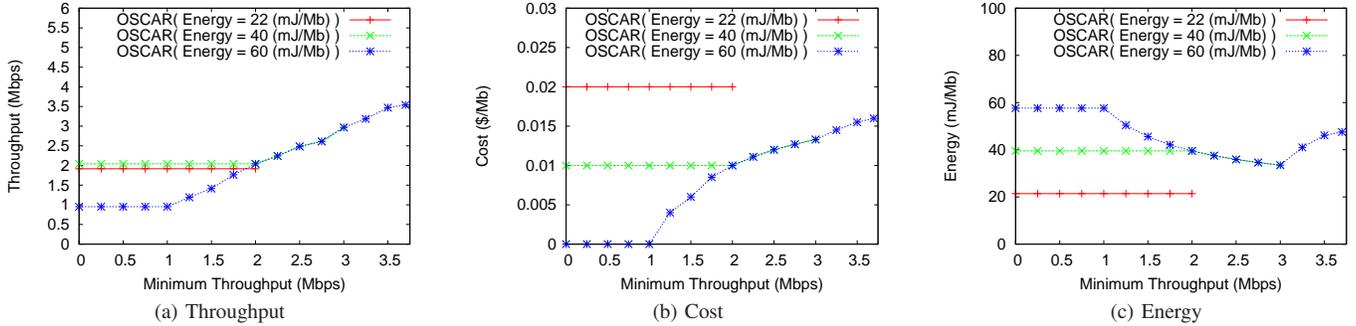

(a) Throughput  (b) Cost  (c) Energy

Fig. 12. Effect of changing the minimum throughput on OSCAR Cost Minimization Mode

Figure 12 shows the effect of changing the user required throughput and the average energy limit on the performance of OSCAR scheduler. In this experiment we first select the throughput constraint and then the energy constraint. Figure 11(b) shows that, as expected, relaxing either energy or throughput constraints leads to less cost consumption per unit data. In the extreme, when the energy limit is relatively high and the required throughput is significantly low, OSCAR assigns all the system traffic to the cost-free path through $IF_1$. Furthermore, OSCAR's average cost per unit data increases with the increased user throughput requirement or the decreased energy limit. On the other hand, the higher throughput requirement puts limits on the energy constraint reduces its the feasibility region.

Figure 12(a) shows OSCAR's ability to meet the user throughput requirement. In addition, It can significantly exceed the user requirements on the throughput as long as this does not neither violate the energy constraint nor increase the average cost per unit data.

Figure 12(c) shows that the relation between user required throughput and average energy consumption per unit data is not straightforward. It depends on the characteristics of the available paths. Hence, we find that the energy consumption remains constant with the increase of the minimum throughput requirement when this increase does not affect the scheduling decision. Then, it decreases with the increase of user throughput requirement when OSCAR starts using a path with lower energy per unit data. After that, It increases with the increase of the user throughput when OSCAR starts utilizing an energy hungry path.

## VIII. CONCLUSION AND FUTURE WORK

We proposed OSCAR, an energy and cost aware, incentive-based, collaborative, and deployable bandwidth aggregation system. We presented the OSCAR architecture, formulated the optimal scheduling problem that aims at maximizing throughput while maintaining the average cost and energy consumption at user defined levels, and presented OSCAR's communication protocols.

We evaluated OSCAR using both implementation and simulation, showing how it can be tuned to achieve different user-defined goals. In the throughput maximization mode, it can provide up to 150% enhancement in throughput compared to current operating systems, without any changes to legacy servers. Moreover, the performance gain further increases with the availability of resume-supporting, or OSCAR-enabled servers, reaching the maximum achievable upper-bound throughput under the conditions of the current Internet.

For future work, we plan to extend our system by optimizing for multiple objectives including trust in addition to throughput, cost, and energy consumption. We also intend to implement our system on mobile operating systems. Finally, we plan to create adaptive scheduling strategies sensitive to user profiles and real-time needs.